\documentclass[a4paper]{jpconf}
\usepackage{graphicx}
\usepackage{here}
\usepackage{siunitx}
\usepackage{url}
\usepackage{wrapfig}
\usepackage[margin=25truemm]{geometry}
\begin{document}
\title{Development of a new high-speed data acquisition system prototype for SOI pixel detector using SiTCP-XG, a 10-gigabit Ethernet network processor}

\vspace*{-0.5\intextsep}

\author{Ryutaro Nishimura$^1$, Shunji Kishimoto$^1$, Yasuo Arai$^1$ and Toshinobu Miyoshi$^1$}

\address{$^1$ High Energy Accelerator Research Organization, Oho 1-1, Tsukuba, Ibaraki, 305-0801, Japan}

\ead{ryunishi@post.kek.jp}

\vspace*{-0.5\intextsep}

\begin{abstract}
We are developing a new readout board with a newer generation field-programmable gate array (FPGA) and the 10-gigabit ethernet to improve the performance and usability of the current readout board based on the 1-gigabit ethernet. 
In this new readout board, the SiTCP-XG network processor supporting 10 Gigabit Ethernet was implemented. 
SiTCP is a network processor circuit running on FPGA, and SiTCP-XG is the newly developed version of the SiTCP that supports 10-gigabit ethernet. 
Before developing the new board, we constructed a prototype system using the Xilinx FPGA evaluation board KC705 to evaluate the SiTCP-XG. 
This prototype system was tested with the SOI pixel detector, which has 425,984 (column 832 \si{\times} row 512 matrix) pixels and a pixel size of 17 \si{\times} 17 \si{\micro\meter^2} at the synchrotron beamlines of the PhotonFactory (KEK).
This was the first test of the X-ray imaging for this system. 
The results showed that this system worked stably with a transfer rate of 682 Mbps (equivalent to a frame rate of 100 fps, limited by detector operation parameters), and also worked stably with a transfer rate of 2.4 Gbps (equivalent to 350 fps, the maximum rate limited by the detector performance). 
These results suggest that the SiTCP-XG system has sufficient transfer performance to cover the SOIPIX detector performance. 
\end{abstract}

\vspace*{-2\intextsep}

\section{Introduction}

Currently, we are developing a new 10-gigabit Ethernet-based data acquisition (DAQ) system for the SOIPIX detector \cite{soi}, using the SiTCP network processor \cite{sitcp} running on a field-programmable gate array (FPGA). 
Details of the SiTCP network processor and the SOIPIX detector are explained below. 

\subsection{SiTCP}

SiTCP is a network processor IP (Intellectual Property) core that can be implemented in an FPGA. 
This technology was originally developed by T. Uchida (KEK ESYS). 
Currently, it is being further developed and supported by Bee Beans Technologies Co., Ltd. 
This network processor has a small circuit size (approximately 3,000 slices), a simple first-in, first-out interface, and a high transfer speed caused by hardware-based implementation. 
In most cases, the signals of the SOIPIX detector are read out by the SiTCP based readout board, named SEABAS2 (Soi EvAluation BoArd with Sitcp 2) \cite{ryunishiD, ryunishi3, ryunishi4}. 
The SEABAS2 system uses the 1-gigabit Ethernet version of SiTCP (conventional 1 GbE SiTCP). 
This board does not have sufficient performance for advanced experiments requiring a high throughput, such as high-frame rate imaging or large-area imaging. 
To solve this issue, we are developing a new readout board consisting of a newer generation FPGA and 10-gigabit Ethernet version of SiTCP. 
This 10-gigabit Ethernet version of SiTCP, named SiTCP-XG, supports 10-gigabit Ethernet while maintaining a compatible interface with conventional SiTCP. 
Thus, the new system achieves ten times higher throughput than conventional SiTCP systems such as SEABAS2. 
As the SiTCP-XG is currently under development, the alpha version released by Bee Beans Technologies was tested using the prototype system described below. 

\subsection{SOIPIX detector}

The SOIPIX detector is a monolithic pixel detector series using silicon-on-insulator technology developed by the SOIPIX group, led by the High Energy Accelerator Research Organization (KEK). 
The SOIPIX detector consists of thick, high-resistivity silicon (Si) substrate for the sensor and a thin Si layer for CMOS circuits. 
This detector is based on a \SI{0.2}{\micro\meter} complementary metal-oxide-semiconductor (CMOS) fully depleted silicon on insulator (FD-SOI) pixel process developed by Lapis Semiconductor Co., Ltd \cite{soi}. 
In this study, the INTPIX4NA integration-type SOIPIX detector is the target of the DAQ system development. 
This detector has 14.1 \si{\times} 8.7 \si{mm^2} sensitive area, 425,984 (column 832 \si{\times} row 512 matrix) pixels, and a pixel size of 17 \si{\times} 17 \si{\micro\meter^2}. 

\section{New high speed data acquisition system}

\begin{wrapfigure}[14]{R}[0mm]{0.54\linewidth}
\centering
\vspace*{-1.2\intextsep}
\includegraphics[width=\linewidth]{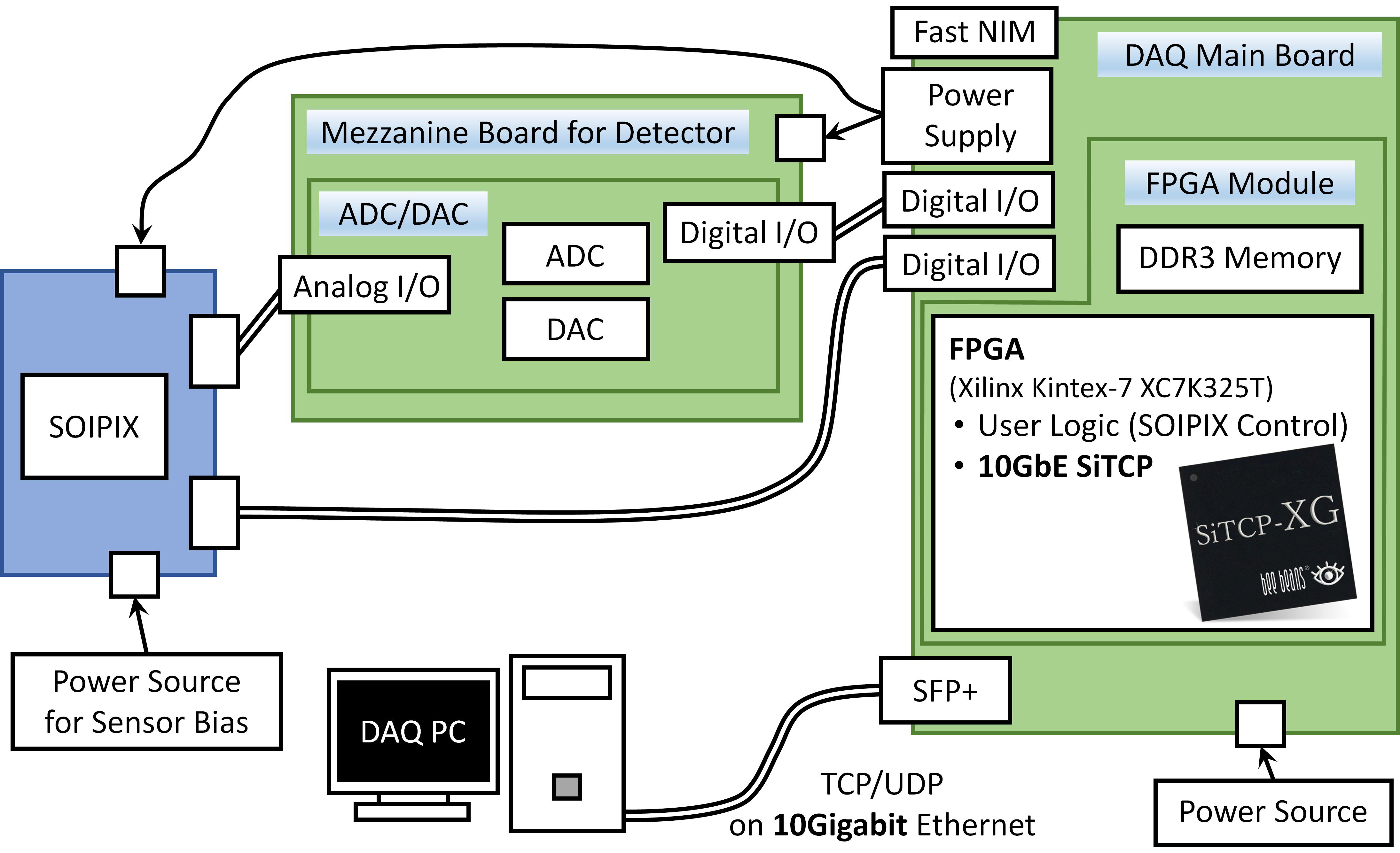}
\vspace*{-1.4\intextsep}
\caption{Schematic of new data acquisition system.}
\label{fig:newboard}
\end{wrapfigure}

A schematic of the new DAQ production board is shown in figure \ref{fig:newboard}. 
System boards consist of a DAQ mainboard (with an FPGA module board) and a mezzanine board for the analog signal processing of the detector.
The mezzanine board has a 16-channel, 12-bit analog-to-digital converter (ADC) and an 8-channel digital-to-analog converter (DAC).
The DAQ mainboard has a digital I/O (MIL-C-83503 40 pin interface) for the SOIPIX detector, a fast NIM interface for external trigger / synchronize control and SFP+ interface for 10-gigabit Ethernet. 
Xilinx Kintex-7 FPGA and DDR3 memory were implemented on a ready-made FPGA module board KX-Card7 \cite{kxcard7} and mounted on the DAQ mainboard. 
The SiTCP-XG was implemented in the Kintex-7 FPGA and mixed implementation with SOIPIX control logic. 
Prior to the development of the new board, a prototype system using the Xilinx commercial FPGA evaluation board KC705 \cite{kc705} was constructed to evaluate the alpha release version of the SiTCP-XG. 

\section{Performance test}

\begin{wrapfigure}[9]{R}[0mm]{0.68\linewidth}
\centering
\vspace*{-2.4\intextsep}
\includegraphics[width=\linewidth]{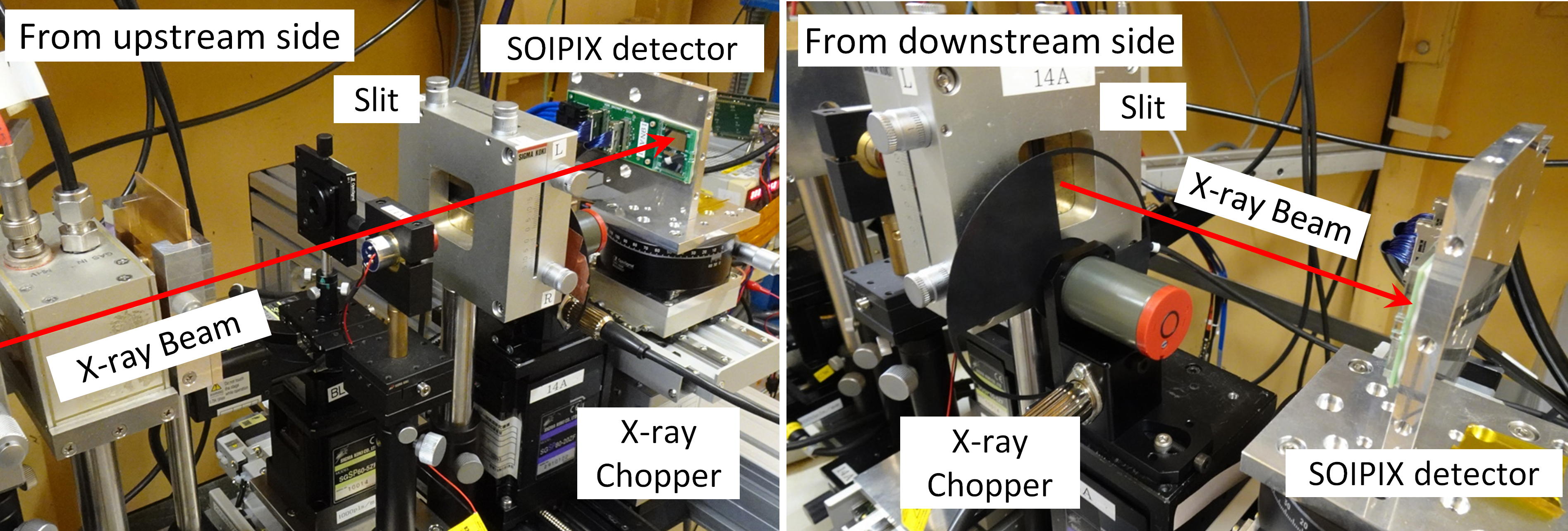}
\vspace*{-1.4\intextsep}
\caption{Test setup photograph from upstream side (left) and downstream side (right). }
\label{fig:setup14a}
\end{wrapfigure}

Tests for the prototype DAQ system were performed at synchrotron beamlines of the PhotonFactory (KEK) BL-14A and BL-14B. 
A \SI{12}{k\electronvolt} X-ray beam was injected from behind the SOIPIX detector and intercepted by an X-ray chopper with a duty ratio of 50 \%. 
Issues in the transfer process of the DAQ presented as anomalies in the X-ray profile. 

This prototype DAQ system was tested with below 3 conditions :
\begin{description}
\setlength{\leftskip}{-15pt}
\setlength{\itemsep}{2pt}
\setlength{\parskip}{0pt}
\setlength{\itemindent}{0pt}
\setlength{\labelsep}{3pt}
\item[Test 1 at BL-14A] Take a 100 fps X-ray image sequence (frame rate was controlled by a 100 Hz internal clock) with the 1 GbE SiTCP or the SiTCP-XG to compare the difference of performance. 
\item[Test 2 at BL-14A] Take a 350 fps X-ray image sequence (frame rate was not controlled by an internal clock) with SiTCP-XG to determine the transfer rate stability with beyond 1 GbE speed. 
\item[Test 3 at BL-14B] Take a 350 fps X-ray image sequence with a larger field X-ray beam. 
\end{description}
In these tests, both the 1 GbE SiTCP system and the SiTCP-XG system were operated on the same KC705 board. 

The test setup at BL-14A is shown in figure \ref{fig:setup14a}. 
The test setup at BL-14B was almost the same except for the X-ray beam size and intensity. 
A \si{\phi}\SI{1}{mm} beam was used in BL-14A, and a rectanglular \SI{10}{mm} (horizontal) \si{\times} \SI{15}{mm} (vertical) beam was used in BL-14B. 

\subsection{Test 1 at BL-14A}

In this test, to check the frame rate stability, the SOIPIX detector's signals are read out with the same parameters via the conventional 1 GbE SiTCP or the SiTCP-XG. 
The detector's exposure time is \SI{200}{\micro\second/frame} \si{\times} \SI{2000}{frames} and the scan time (the settling time for analog output) is 240 \si{ns/pixel}. 
The frame rate is set at 100 fps on configuration, equivalent to 682 Mbps of average data traffic. 
The stability of the frame rate is evaluated using the transition of the interframe period, the time interval between the start of exposure of one frame and the start of exposure of the next frame. 
If the frame rate is completely stable, the interframe period is 10 ms. 
However, if the frame rate is not stable because of the instability of the transfer rate caused by issues in the transferring process, some interframe periods will extend to \SI{20}{ms} or \SI{30}{ms}. 

\begin{figure}[H]
\centering
\begin{minipage}[b]{0.485\linewidth}
\centering
\vspace*{-0.8\intextsep}
\includegraphics[width=\linewidth]{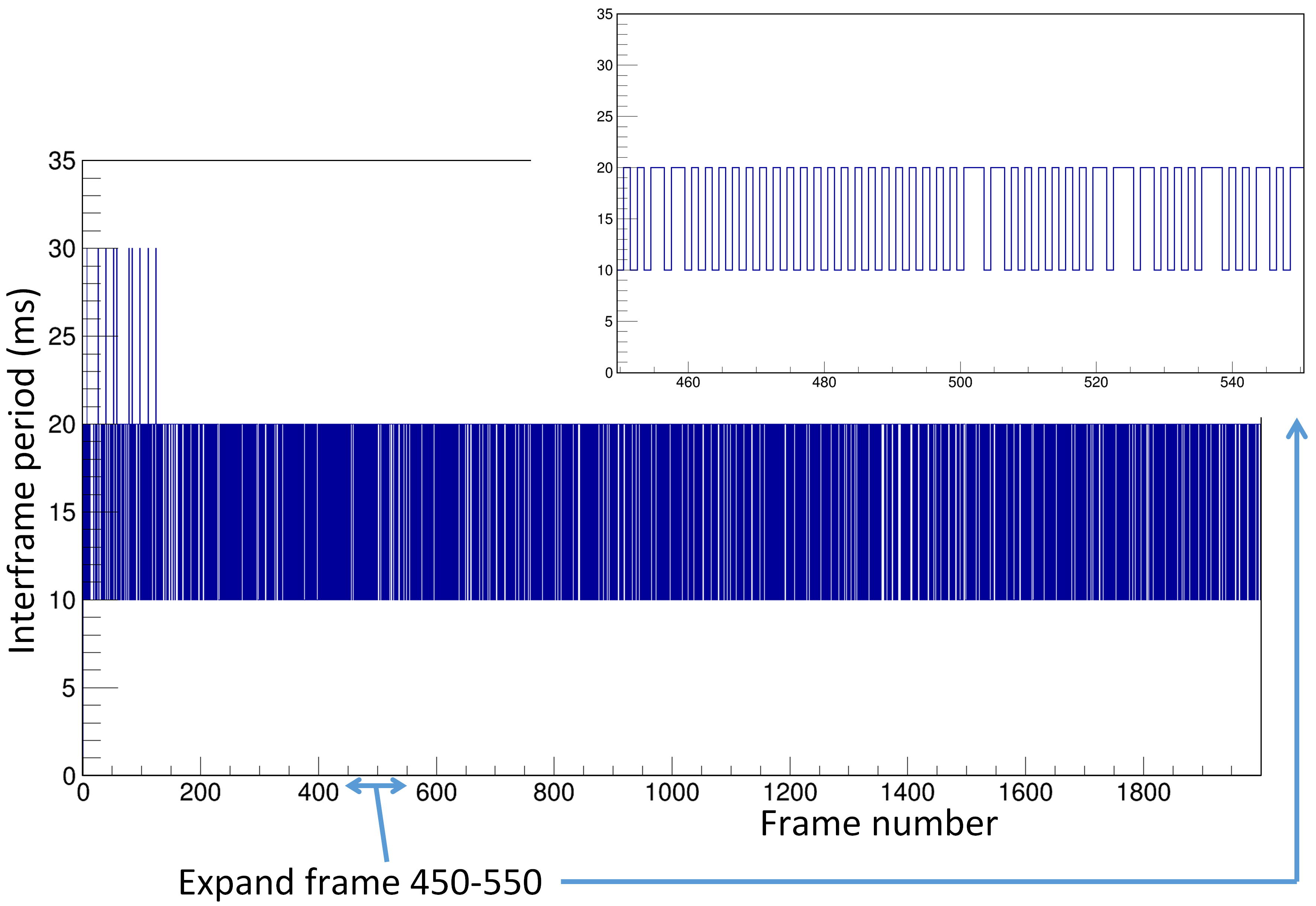}
\vspace*{-1.2\intextsep}
\caption{Transition of interframe period taken by 1 GbE SiTCP system in test 1. }
\label{fig:test1result1g}
\end{minipage}
\begin{minipage}[b]{0.485\linewidth}
\centering
\vspace*{-0.8\intextsep}
\includegraphics[width=\linewidth]{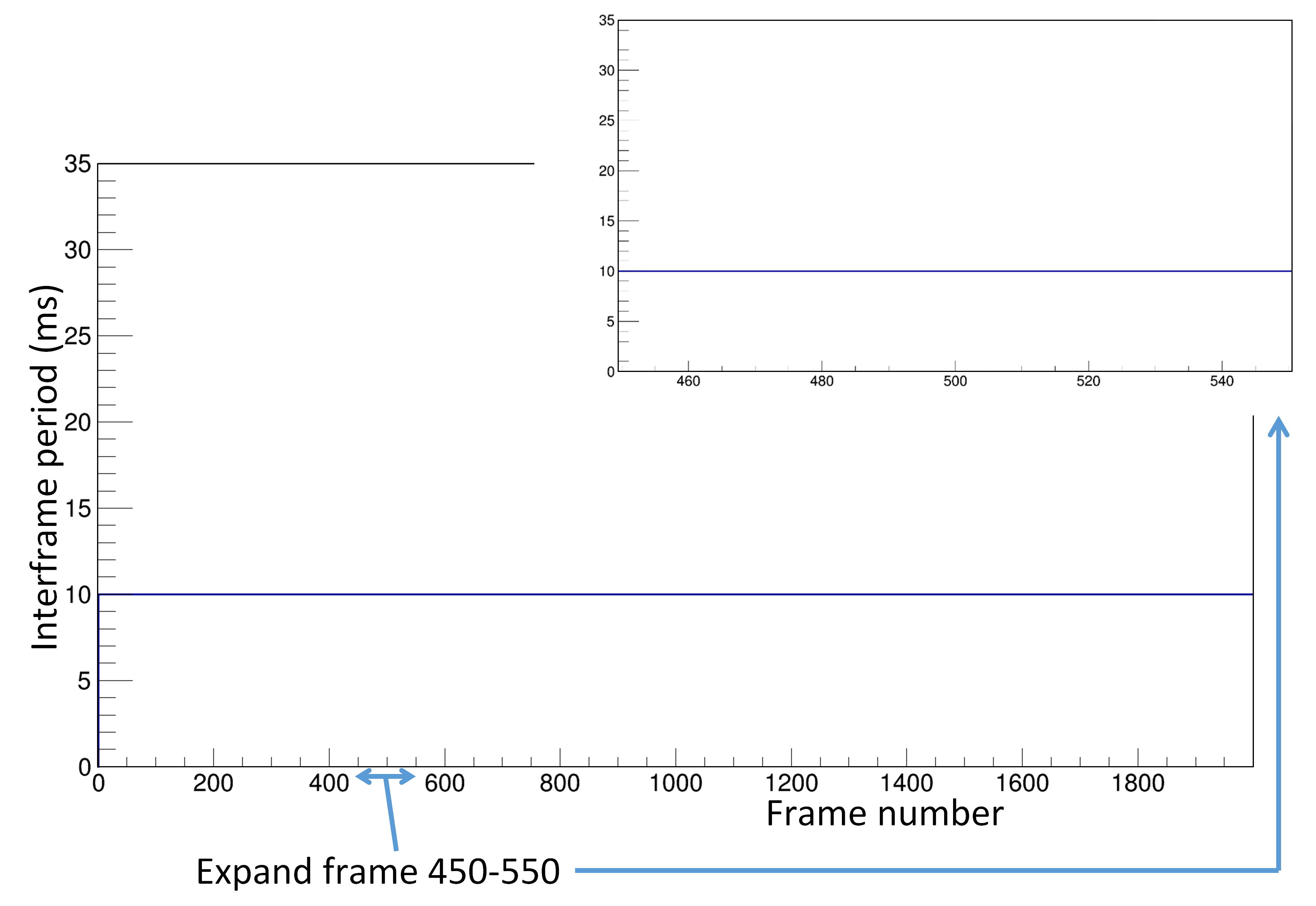}
\vspace*{-1.2\intextsep}
\caption{Transition of interframe period taken by SiTCP-XG system in test 1. }
\label{fig:test1result10g}
\end{minipage}
\end{figure}
\vspace*{-0.8\intextsep}

The results are shown in figure \ref{fig:test1result1g} and \ref{fig:test1result10g}. 
These are the transition plots of the interframe period. 
Figure \ref{fig:test1result1g} are the results of the 1 GbE SiTCP system and figure \ref{fig:test1result10g} are the results of the SiTCP-XG system. 
In the 1 GbE SiTCP result, a few points of the interframe period show \SI{20}{ms} or \SI{30}{ms} values. 
These results indicate that the 1 GbE SiTCP system is not stable at a transfer rate of 682 Mbps because of issues in the transfer process. 
On the other hand, in the SiTCP-XG result, no points show \SI{20}{ms} or \SI{30}{ms} values and a plateau at \SI{10}{ms}. 
Hence, the SiTCP-XG system is stable at a transfer rate of 682 Mbps. 

\subsection{Test 2 at BL-14A}

\begin{wrapfigure}[12]{R}[0mm]{0.52\linewidth}
\centering
\vspace*{-2.4\intextsep}
\includegraphics[width=\linewidth]{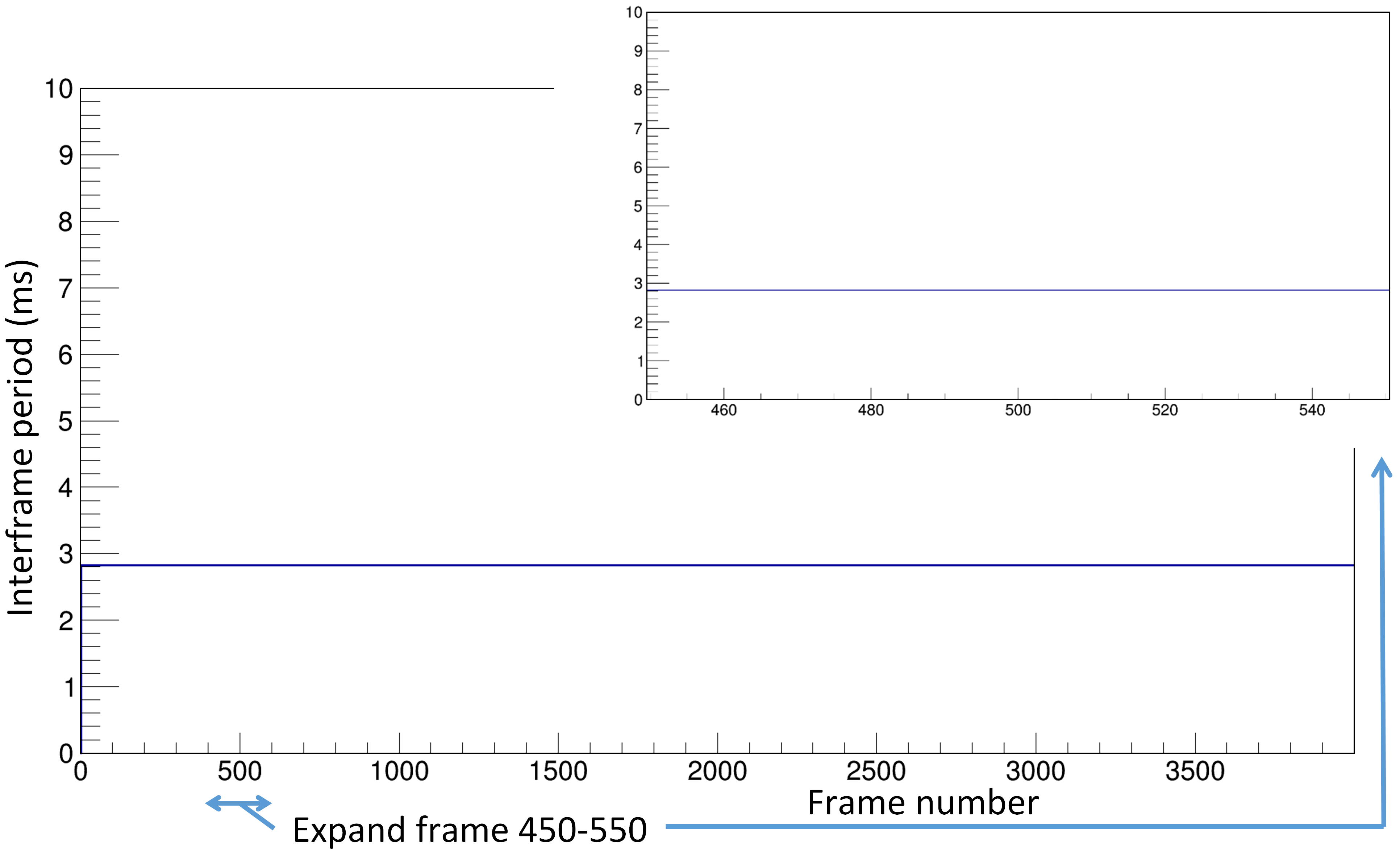}
\vspace*{-1.4\intextsep}
\caption{Transition of interframe period taken by SiTCP-XG system in test 2. }
\label{fig:test2result10g}
\end{wrapfigure}

In this test, to check the stability of the frame rate, the SOIPIX detector's signals are read out with the high-speed readout parameters via the SiTCP-XG. 
The detector's exposure time is \SI{200}{\micro\second/frame} \si{\times} 4000 frames and the scan time is 80 \si{ns/pixel}. 
In this readout condition, the SOIPIX detector's signals are read out with no frame control, and the imaging process is repeated as fast as possible. 
The estimated frame rate is approximately 350 fps, equivalent to 2.4 Gbps of average data traffic. 
The stability of the frame rate is evaluated using the transition of the interframe period. 
If the frame rate is completely stable, the interframe period is \SI{2.828}{ms}. 

Figure \ref{fig:test2result10g} shows the results of the SiTCP-XG system. 
In this system, all points plateaued at \SI{2.828}{ms}. 
Hence, we can conclude that the SiTCP-XG system is stable with a 2.4 Gbps transfer rate and has adequate transfer performance comparable to the SOIPIX detector. 

\subsection{Test 3 at BL-14B}

\begin{wrapfigure}[12]{R}[0mm]{0.5\linewidth}
\centering
\vspace*{-2\intextsep}
\includegraphics[width=\linewidth]{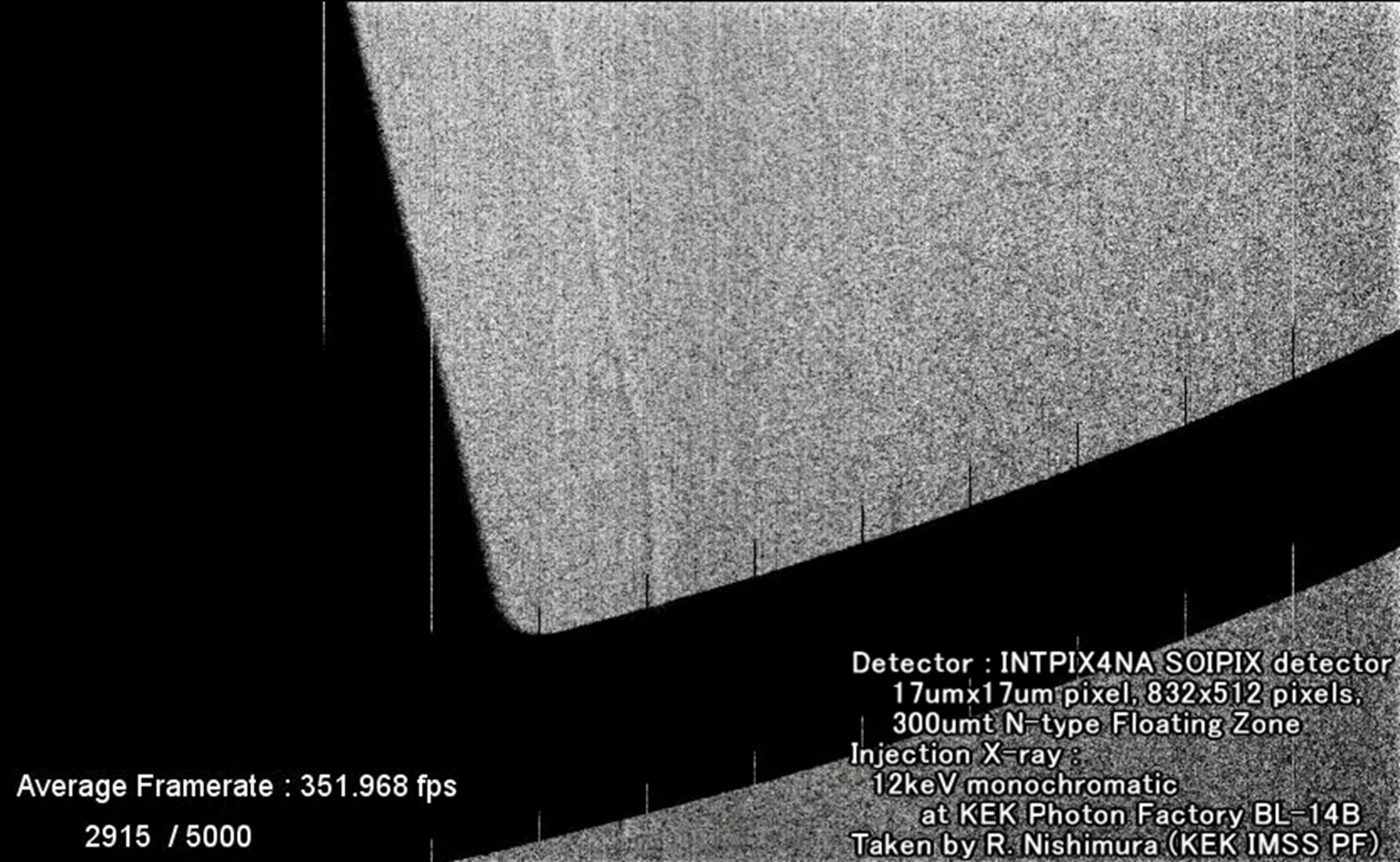}
\vspace*{-1.4\intextsep}
\caption{Capture of X-ray image sequence taken by SiTCP-XG system in test 3. }
\label{fig:test3result10g}
\end{wrapfigure}

This test was performed to demonstrate the performance of the SiTCP-XG system at BL-14B. 
The parameters are almost the same as those in Test 2 without frame number (\SI{200}{\micro\second/frame} \si{\times} 5000 frames), and a rectangular beam of \SI{10}{mm} (horizontal) \si{\times} \SI{15}{mm} (vertical) was used. 

Figure \ref{fig:test3result10g} is the capture image of the X-ray image sequence of rotating X-ray chopper taken at \SI{350}{fps}. 
Full image sequence data are shown on KEK wiki web page \cite{350fpsmov}. 

\section{Conclusion}

The new DAQ system based on a newer generation FPGA and the 10-gigabit Ethernet is currently under development to improve the performance and usability of the SOIPIX detector. 
The SiTCP-XG, the newly developed version of the FPGA-based network processor supporting 10-gigabit Ethernet, will be implemented in this new system. 
This SiTCP-XG is currently under development; thus, the alpha version was tested with the prototype of the DAQ system constructed on the KC705 FPGA board to evaluate its performance. 
This prototype system was tested with an SOI pixel detector, which has 425,984 (column 832 \si{\times} row 512 matrix) pixels, at synchrotron beamlines of the PhotonFactory (KEK).
The results show that the SiTCP-XG system is stable with a transfer rate of 682 Mbps (equivalent to a frame rate of 100 fps), and also is stable with a transfer rate of 2.4 Gbps (equivalent to 350 fps). 
Hence, we can infer that the SiTCP-XG system has an adequate transfer performance comparable to the SOIPIX detector. 
These are the initial results of the X-ray imaging data taken by the SiTCP-XG-based system. 

\section*{Acknowledgments}

This study was conducted under the approval of the Photon Factory Program Advisory Committee (Proposal No. 2020PF-21 and 2020PF-33). 
This study was performed on behalf of SOIPIX collaboration \cite{soipix}. 
The alpha release version of SiTCP-XG was provided by Bee Beans Technologies Co., Ltd.

\end{document}